# Direct X-ray detection of the spin Hall effect in CuBi

Sandra Ruiz-Gómez[1,2], Rubén Guerrero[3], Muhammad W. Khaliq[2], Claudia Fernández-González[1,3], Jordi Prat[2], Andrés Valera[3], Simone Finizio[4], Paolo Perna[3], Julio Camarero[3,5], Lucas Pérez[1,3,6], Lucía Aballe[2], Michael Foerster[2]

[1] Departamento de Física de Materiales, Universidad Complutense de Madrid, Plaza de las Ciencias 1, 28040 Madrid, Spain

[2] ALBA Synchrotron Light Facility, CELLS, Carrer de la Llum, 2-26, E-08290 Bellaterra, Spain

[3] Instituto Madrileño de Estudios Avanzados, IMDEA Nanociencia, Calle Faraday 9, 28049 Madrid, Spain

[4] Swiss Light Source, Paul Scherrer Institut, CH-5232 Villigen PSI, Switzerland

[5] Departamento de Física de la Materia Condensada e Instituto "Nicolás Cabrera" and Condensed Matter Physics Center (IFIMAC), Universidad Autónoma de Madrid (UAM), Campus de Cantoblanco. Madrid 28049, Spain

[6] Surface Science and Magnetism of Low Dimensional Systems, UCM, Unidad Asociada al CSIC (IQFR)

**Abstract:** The spin Hall effect and its inverse are important spin-charge conversion mechanisms. The direct spin Hall effect induces a surface spin accumulation from a transverse charge current due to spin orbit coupling even in non-magnetic conductors. However, most detection schemes involve additional interfaces, leading to large scattering in reported data. Here we perform interface free x-ray spectroscopy measurements at the Cu $L_{3,2}$ absorption edges of highly Bi-doped Cu ($Cu_{95}Bi_5$). The detected X-ray magnetic circular dichroism (XMCD) signal corresponds to an induced magnetic moment of $(2.7 \pm 0.5) \times 10^{-12} \mu_B\ A^{-1}\ cm^2$ per Cu atom averaged over the probing depth, which is of the same order as for Pt measured by magneto-optics. The results highlight the importance of interface free measurements to assess material parameters and the potential of CuBi for spin-charge conversion applications.

**Introduction**

Since the discovery of the giant magnetoresistance effect triggered the field of spintronics, spin dependent electron transport phenomena have been widely investigated and gained importance in research and technological applications. Within spintronics, the spin Hall effect (SHE) may be considered a more recent discovery [1], although it had been predicted several decades ago (Dyakonov and Perel [2], based on Mott scattering [3]). The SHE and its inverse refer to orthogonal charge and spin currents which can arise through different mechanisms, but always due to spin orbit coupling. For comprehensive reviews please refer to Ref. [4, 5]. The SHE is already widely used to generate and detect spin currents. It can drive magnetic excitations more efficiently than spin transfer torque (STT), for instance in spin orbit torque magnetic random access memory (SOT-MRAM). The direct SHE generates a pure spin current in a non-magnetic conductor, perpendicular to the electron flow, with the spin polarization perpendicular to both. The ratio of spin and charge current is the spin Hall angle (SHA) $\alpha = \frac{\sigma_{xy}^S}{\sigma_{xx}^C}\frac{e}{\hbar}$ [5], where $\sigma_{xy}^S$ and $\sigma_{xx}^C$ are the spin and charge current conductivities, respectively. The SHA measures the efficiency of charge to spin conversion and is thus the main figure of merit for applications. The spin current results in a spin accumulation at the edges of the conductor, on the length scale of the spin diffusion length (see scheme in Figure 1a).

First observations of the SHE in semiconductors used optical detection [6, 7], but optical methods have proven challenging for metallic systems, due to their considerably shorter spin diffusion lengths. In metals, electrical detection schemes like non-local spin valves, inverse spin Hall effect, spin Seebeck effect, spin orbit torque, etc are typically used. However, all those techniques involve an interface with another magnetic material, and as a consequence, the combined materials' and interface properties are measured, which might include other sources of spin orbit interactions such as the Rashba effect. Possibly this is one of the reasons for the variability of reported results for the spin Hall angle and diffusion length/lifetime in metals. The large scattering of reported results from these methods for the same material and temperature (e.g. see table III of [4]) call out for a more direct and interface-free approach. Only recently, optical measurements for Pt and W have been reported [8].

X-ray spectroscopy and magnetic circular dichroism (XMCD) has become a reference tool for precision measurements of small or diluted magnetic signals [9-13]. X-ray detection of the SHE is highly desirable since it gives element specific access to the electronic information and can provide a clear spectroscopic fingerprint. Furthermore, performing the measurement on a single layer can eliminate the influence of an interface. Still using heterostructures, Stamm et al. [14] investigated with XMCD spectroscopy Pt/Co and Pt/NM (NM=Ti,Cr,Cu) bilayers for a spin accumulation in the overlayer due to the SHE in Pt. While a clear change in the Co moment was observed, no signal above noise limit was detected in the NM for the other cases. Since the expected spin accumulation in Pt should have been above the detection limit, the lack of a clear signal may again be due to the presence of the Pt/NM interface. Thus, to the best of our knowledge, the direct detection of spin accumulation due to the SHE by X-ray spectroscopic methods is still lacking.

As a general trend, heavy metals such as Pt, W, etc. with high spin-orbit coupling, are expected to give rise to comparatively large spin Hall angles (SHA). Although Cu does not exhibit a high SHA, it has been experimentally demonstrated that by adding only 0.5% Bi doping, a giant SHA $\alpha_{CuBi}$ = -0.11 (for the CuBi alloy and $\alpha_{imp}$ = -0.24 for the skew scattering mechanism) [15] can be obtained, higher than the one of Pt $\alpha_{Pt}$ = 0.068 [16, 17] ($\alpha_{Pt}$ = 0.08 in [8]). More recently, thin Cu films with higher Bi doping levels, up to 10%, have been prepared, without signs of Bi segregation [18]. Interfaces of these films with Yttrium Iron Garnet (YIG) exhibit a large spin mixing conductance $g_{\uparrow\downarrow}$ [19, 20], indicated by the increased damping in spin pumping ferromagnetic resonance (FMR) measurements, yielding values $g_{\uparrow\downarrow}$ ($Cu_{96}Bi_4$/YIG) = 7 x $10^{18}$ m$^{-2}$) [18] comparable to those of Pt/YIG ($g_{\uparrow\downarrow}$ (Pt/YIG) = 6.9 to 9 x $10^{18}$ m$^{-2}$) [19]). However, based on the work of Niimi et al [15] for CuBi, one may have anticipated an even higher SHE in such highly Bi-doped Cu. This discrepancy could be related to the quality of the Pt/YIG and CuBi/YIG interfaces, highlighting again the need for a more direct method to assess true material parameters.

CuBi is a good candidate for the X-ray detection of the SHE due to the convenient Cu-$L_{3,2}$ absorption edge in the soft X-ray range. In fact, Kukreja et al [21] reported soft X-ray spectroscopic measurements of a transient spin accumulation in Cu induced by a spin polarized current from an adjacent Co layer, proving that small spin accumulations in Cu can be indeed be measured with X-rays. They identified a spectral feature of the spin accumulation at the rising edge of the $L_3$ absorption peak, corresponding to transitions to the Fermi level. In addition, a Cu XMCD signal induced by proximity to Co and enhanced by current injection was found close to the $L_3$ peak. More recently, using the same setup, Ding et al. have observed a spin accumulation in Cu induced by spin pumping from an adjacent $Ni_{80}Fe_{20}$ layer detected at the Cu $L_3$ edge [22].

Here we report the detection of a spin accumulation under electric current flow, i.e. spin Hall effect, in the surface of highly Bi-doped $Cu_{95}Bi_5$ using soft X-ray photoemission electron microscopy (PEEM). The measurements at the $L_{3,2}$ absorption edges were performed on the top of single material CuBi electrodes on insulating $SiO_x$/Si substrates, i.e. free of any interface effect. The PEEM information depth is limited to about 5 nm by the electron escape length, allowing us to selectively detect the spin accumulation in the upper surface of the electrode (the sensitivity as function of depth z is approximately $e^{-z/2 \text{ nm}}$ [23]). In other words, the signals from upper and lower surface do not compensate in PEEM as they would if the whole thickness was measured, e.g. in transmission geometry. The microscopy approach enables us to focus on small electrode structures and achieve high current densities at moderate total power dissipated. The sample geometry was chosen to compensate different backgrounds by the use of a symmetric electrode design (Figure 1). This layout enables us to reduce the background variations, in particular those associated to the voltage drop along the electrode due to the current and to the inhomogeneity of the X- ray illumination for opposite polarization. Thanks to these factors, we have been able to detect a clear dichroic (XMCD) signal of the X-ray absorption in CuBi associated to a SHE-induced surface spin accumulation.

**Experimental:**

$Cu_{95}Bi_5$ thin films were deposited by co-evaporation at room temperature onto Si substrates with a 200 nm thick thermally grown oxide layer to avoid current shunting in the experiments. The electrode structures were patterned by electron beam lithography followed by lift-off (details in Appendix A). The preparation and characterization of highly Bi doped Cu films have been described in Ref. [18]. The thickness of the CuBi electrodes was varied between 20-50 nm. After growth, samples were capped in-situ with a 3 nm $Al_2O_3$ layer in order to avoid surface oxidation.

X-ray spectroscopic measurements were performed at the LEEM/PEEM experimental station of the CIRCE beamline of the ALBA Synchrotron, which has excellent stability [24]. Samples were mounted onto printed circuit boards (PCBs) and contacted with wirebonds, using dedicated sample holders [25]. For most data, direct current (DC, 30-120 mA) was injected into the electrodes using a current source purposely designed for the PEEM instrument. A single data set was obtained using pulsed current injection (millisecond range) to reduce heating and allow even higher current density. During measurements, the sample holder was cooled down to about 220 K. The DC currents of 30-120 mA injected into the samples resulted in a moderate increase of the pressure in the UHV chamber from Joule heating (reaching $10^{-9}$ mbar while starting in the $10^{-10}$ mbar range). For too high current values, we observed in real time the segregation of small clusters to the surface, which are rich in Bi (see Appendix B). All the data points used in this study were acquired below such current limit. In the PPEM instrument, the X-ray beam incidence onto the sample is 16° grazing, which results in a mixed sensitivity to in-plane and out-of-plane magnetic moments in XMCD measurements (with 0.96 and 0.28 efficiency, respectively). Therefore, assuming a pure in-plane spin polarization in our experiment, a correction factor of 1.04 is applied to the final extracted XMCD values. To obtain a basic measurement, i.e. an XMCD image at a fixed photon energy, we averaged a number (typically 100-200) of single images for each of the two circular polarizations (left (CP) and right-handed (CN)) each, before taking the pixelwise asymmetry XMCD = {I(CP)-I(CN)}/{I(CP)+I(CN)}, where I(CP/CN) denotes the countrate of the low energy secondary electrons for a pixel. The XMCD value for one part of the electrode is then taken as the average of all pixels in that area.

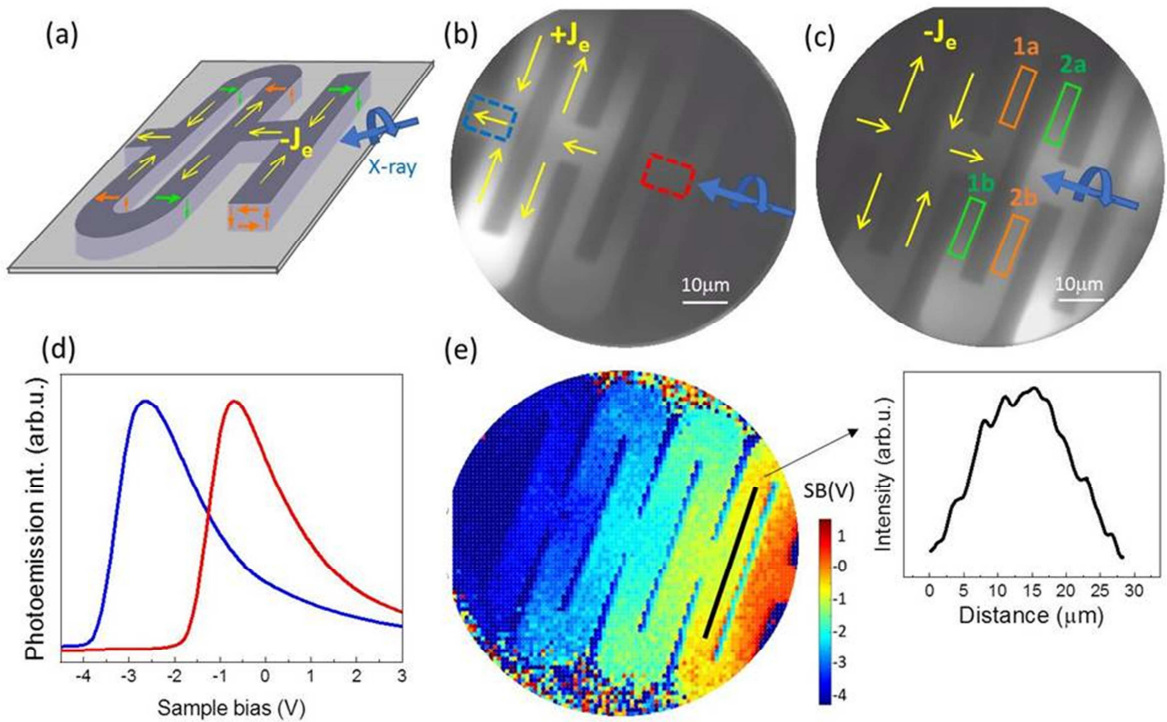

*Figure 1: Sample layout and effect of current injection. (a) Schematic of the experimental geometry, indicating the electron flow (yellow arrows), as well as the accumulated spin polarization at the different surfaces (green and orange arrows). The circularly polarized X-ray beam, shown in blue, impinges from the right under 16° grazing angle, providing sensitivity to the spin polarization in the beam direction. (b) PEEM image of a measured device during positive electron flow (negative current) (yellow arrows). The surface electric potential variation due to the driving voltage causes an intensity variation from left to right. Red and blue boxes show the areas corresponding to the spectra in panel (d). (c) PEEM image for negative electron flow. The associated potential drop is opposite than in panel (b) The green and orange boxes in (c) indicate the two types of areas with the same current direction and thus spin accumulation. (d) Spectra of the photoemission intensity vs sample bias voltage for the blue and red boxes in panel (b). The different local potential is reflected in a shift of the low energy secondary electron peak vs sample bias voltage. (e) Color map of the sample bias voltage value at which the maximum of photoemission intensity is located. The profile along the black line (inset) indicates that the current is correctly flowing through the electrodes.*

**Results and discussion**

The schematic of the experiment is shown in Figure 1(a), where the circularly polarized x-ray beam illuminates the sample from the right (blue arrow). The yellow arrows mark the direction of the electron flow in the electrode loops, being opposite in panels (b) and (c). The current injection (current density j = 1.7 x $10^7$ Acm$^{-2}$) gives rise to a voltage drop along the path of the electrons, which is detected in PEEM as a shift of the secondary electron spectrum with respect to the sample bias voltage (nominal photoelectron kinetic energy). In panel (d) we show two spectra obtained from the red and blue areas of panel (b), with a relative shift of about 2 V. The color map in panel (e) indicates at which bias voltage the spectra reached maximum intensity for a given area. The potential variation along the black line as shown in the inset in panel (e) demonstrates that the current is indeed following the electrode path. The potential variation along the electrode is reproduced by finite element simulations and indicates a resistivity of 13 µΩcm of the patterned CuBi electrode at 220 K (see Appendix C). Images in panels (b) and (c) were taken at constant sample bias voltage and thus reflect the potential variation as intensity. The intensity variation will differ for opposite driving current, making a measurement scheme with alternating injected currents (or lock-in detection) impractical for PEEM. Therefore, the electrode layout was chosen to provide equipotential points in the middle, permitting the direct comparison of areas with the same electric potential but opposite current direction. Those areas are vertically aligned, for example the orange and green boxes, labelled 1a and 1b, in Figure 1 (c). Since the accumulated spin polarization is opposite in consecutive branches (e.g. 1 and 2), the dichroic component of the X-ray absorption, taken as 1/2 of the up-down asymmetry for each branch, is alternating.

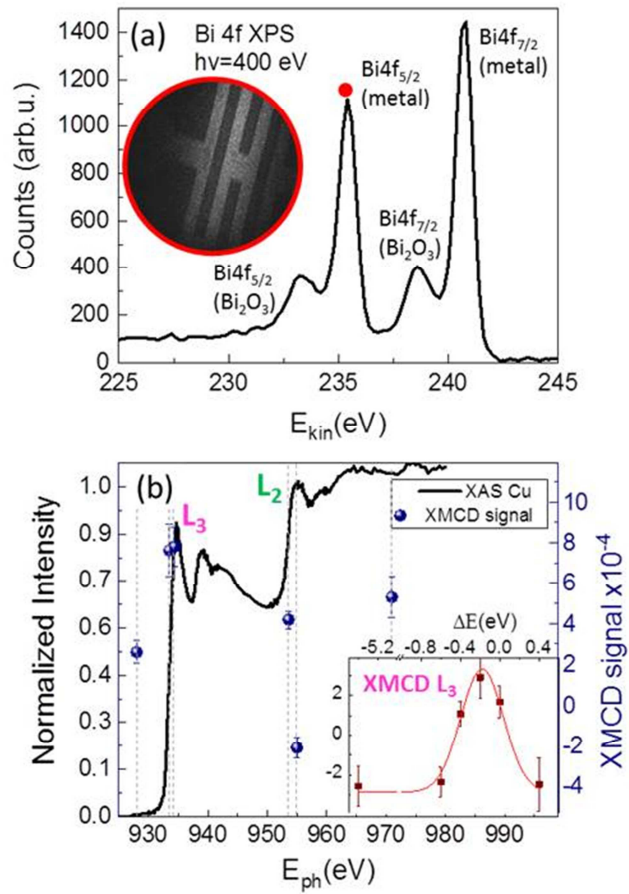

Figure 2: X-ray spectroscopy of the CuBi electrodes and the spin accumulation signal. (a) Photoelectron spectrum of the Bi 4f core level, showing a small surface oxide contribution. Inset: XPEEM image at the Bi4f $_{5/2}$ peak (marked with a red dot in the spectrum). Bismuth is homogeneously distributed in the electrode surface. (b) X-ray absorption

*spectrum of the Cu L edges together with one data set of XMCD values. Inset: XMCD scan close to the $L_3$ edge for a different sample (Y-axis shows the same units as the left axis of the main panel).*

To characterize the $Cu_{95}Bi_5$ electrodes, we have performed X-ray photoelectron spectroscopy (XPS) and X-ray absorption spectroscopy (XAS). The protective Al capping layer was removed in-situ by Ar ion sputtering before all measurements. The Bi 4f XPS shows mostly metallic Bi with a small oxide component (Figure 2(a)). In contrast, measurements of samples exposed to air after the capping removal showed almost fully oxidized Bi (Appendix D)). XPS spectra are extremely surface sensitive, with a probing depth in the order of 1 nm. The XAS of the Cu $L_{3,2}$ edges indicates that the Cu in the electrode is also metallic (Figure 2(b)). Together with the spectrum, the extracted XMCD signal is plotted for different photon energies. Each single data point is the average of at least 3 independent measurements. The method for the determination of the XMCD signal, which involves a fit of the background, is explained in Appendix E.

As can be seen in Figure 2 (b), the SHE induced XMCD signal has a maximum at the $L_3$ edge and is inverted in the $L_2$ edge. However, even for off-peak energies, where no XMCD signal is expected, finite values are found (see also Figure 9). This offset was found to be inconsistent between different measurements series and thus we have to conclude that even after best efforts, tiny background variations in the XMCD images are of the same order of magnitude as the SHE signal in our data. These variations are most likely coming from the different beam intensity distribution for opposite beam polarization. However, for each single run at a fixed current value, after thermalization, the offset was roughly constant over time and photon energy, so that we can normalize the XMCD signal for the $L_3$ and $L_2$ peaks by subtracting the value obtained by an immediately preceding measurement at a pre-peak photon energy. A similar offset (of opposite sign) is seen in the inset of Figure 2(b) where a scan of the XMCD signal close to the $L_3$ absorption edge is shown for a different sample. A Gaussian fit to the XMCD data yields a FWHM of 0.47 eV and peak position at -0.2 eV with respect to the $L_3$ peak. Note that both features, the narrow peak width (close to the minimum allowed due to the core hole lifetime) and the maximum location before the XAS peak, are characteristic of a spin accumulation at the Fermi edge [21].

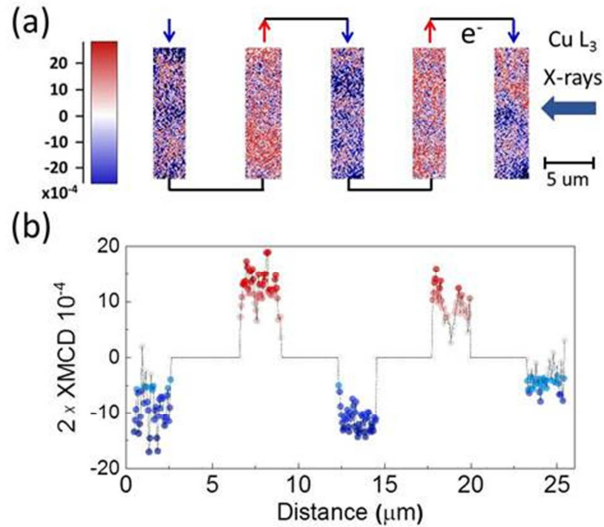

*Figure 3: (a) Visualization of the spin accumulation in a $Cu_{95}Bi_5$ electrode. The graph shows the pixelwise up-down asymmetry of an XMCD image taken at the Cu $L_3$ edge, i.e. the subtraction of two areas with opposite current flow. Areas outside of the integration regions have been masked in order to improve clarity. (b) Horizontal profile through the data of panel (a), highlighting the alternating signal.*

In Figure 3(a), we show a visualization of the spin accumulation in the $Cu_{95}Bi_5$ electrode by taking the pixelwise up-down asymmetry, i.e. the difference of the XMCD signal from the upper and lower part of each electrode branch, for one of the images acquired at the Cu $L_3$ edge. To improve the clarity, the areas outside the integration regions (i.e., outside the center of the electrodes) have been masked and the background flattened. The single XMCD image was taken from one run where the offset was small, and no pre-peak subtraction was performed here. A clear *low-high-low-high-low* (blue-red-blue-red-blue) pattern in the five consecutive electrode branches can be observed, reflecting the alternating spin accumulation due to the alternating current direction. This oscillatory pattern is also visible in the profile of Figure 3(b) which averages data of panel (a) in the vertical direction. The full average over each area is the input of the fits which are used to quantify the XMCD signal (see Appendix E)).

Figure 4 summarizes all the XMCD signals ($Cu_{95}Bi_5$ thickness between 20 and 50 nm, as indicated) where pre-peak measurements were available to perform the subtraction. Datasets without pre-peak normalization have been discarded, although some of them appeared reasonable (see for example Figure 10), for the sake of consistency throughout the analysis. The thickness of all samples reported in Figure 4 is well above the information depth and thus the exclusive probing of the upper surface is guaranteed. Each data point is the average of at least 3 single measurements for each energy, peak and pre-peak, respectively. The error bars represent the combined errors of the two photon energies. The data shows a consistent sign inversion with respect to the injected current and the absorption peak ($L_3$ vs $L_2$). Dashed lines are linear fits to the data which yield XMCD coefficients of $(10 \pm 2) \times 10^{-12}$ cm$^2$ A$^{-1}$ for the Cu L$_3$ and $(-9 \pm 1) \times 10^{-12}$ cm$^2$ A$^{-1}$ for the Cu L$_2$.

Note that the absolute sign of the XMCD effect at the Cu L$_{3,2}$ absorption peaks due to spin polarization at the Fermi level should be the same as for the 3d magnetic metals such as Fe, because in both cases more empty *minority* final states are available for transitions. The measured sign is in agreement with a negative spin Hall angle in CuBi [15], i.e., when the electron flow is upwards in Figures 1 (b,c), a spin polarization parallel to the incident beam (blue arrow) is produced at the upper surface of the electrode.

The measured XMCD signals due to SHE at the Cu L$_3$ and L$_2$, as given by the linear fits of Figure 4, i.e. $(10 \pm 2) \times 10^{-12}$ cm$^2$ A$^{-1}$ and $(-9 \pm 1) \times 10^{-12}$ cm$^2$ A$^{-1}$, are of opposite sign but almost equal strength. This relation is in agreement with Ref. [26], reporting XMCD spectra for a proximity induced magnetization at the Cu d shell for thin Cu layers sandwiched with Co, due to the hybridization between Co and Cu d orbitals. The reason a comparison of that data with the SHE induced spin polarization can be attempted is that even in the absence of Co, i.e. for pure bulk fcc Cu, the L$_{3,2}$ edges are probing almost exclusively 2p-3d transitions, due to the 3d-4s hybridization at the Fermi level and very different transition matrix elements [27]. The induced magnetic moment is determined using the same procedure as in the case of spin polarized current injection from a Co electrode, reported by Kukreja et al. [21] who also consider transitions into 3d states using references [26, 27]. The estimation is a lower bound because the XMCD values were measured at the XAS peak, located 0.2 eV above the peak of the XMCD spectrum. Using Ref [26] and the width of a Gaussian XMCD peak profile in Figure 2(b), 0.47 eV, the XMCD coefficient at the Cu L$_3$ translates to an induced magnetic moment of $(2.7 \pm 0.5) \times 10^{-12}$ $\mu_B$ cm$^2$ A$^{-1}$ per Cu atom as weighted average over the information depth (around 5 nm). For our typical current densities the average detected magnetic moment per atom is thus about $7 \times 10^{-5}$ $\mu_B$/atom.

For comparison, the paramagnetic spin polarization due to the current induced Oersted field is below $1 \times 10^{-6}$ $\mu_B$/atom (see SI 6). Concerning sample imperfections as possible spurious origins of the measured XMCD signal, the edges of the electrodes were excluded from the analysis and we confirmed that when shifting the areas of integration the resulting numbers did not show relevant changes. Also since similar results were obtained on four different samples, local defects from lithography or sample homogeneity can be ruled out. Neither the observed background variations in the XMCD images, caused by the beam intensity distribution and the sensitivity to the local potential from the driving voltage, can explain the measured signal, although they are assumed to be the origin for the offset, for example the shift towards positive XMCD values in Figure 2(b). The beam intensity distribution does not change with current polarity, and therefore at most could produce a random sign, but not a consistent sign change for opposite current directions. On the other hand, a possible artefact originating from the local potential would change sign with the current direction. That is the reason why a direct comparison of images with opposite current directions or a lock-in detection scheme are not feasible in the PEEM experiment. Instead, as mentioned previously, we achieved the direct comparison for opposite current directions by comparing the upper and lower part of the electrode, which are joined by equipotential points (virtual ground) in the middle. However, the effect of the local potential does not depend on the photon energy of the incoming X-ray beam and therefore it is eliminated by the subtraction of the offset from pre-edge energies. Also it could not induce a peak shape of the XMCD signal as observed in the inset of Figure 2(b). Finally, any hypothetical artefact related to a combination of a previously mentioned effect together with sample charging, which could change at an absorption peak, cannot reproduce the inversion of the signal between the $L_3$ and $L_2$ peaks.

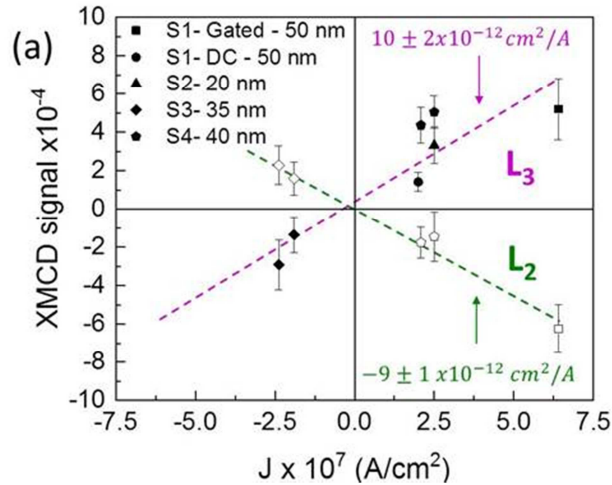

Figure 4: XMCD signal as function of current density for measurements at the Cu $L_3$ and $L_2$ absorption peak energies, extracted as explained in the main text. Each point is an average of at least 3 measurements. The error bars correspond to the combined errors of peak and pre-peak values. The dashed lines are linear fits to the data.

The determined magnetic moment in $Cu_{95}Bi_5$, $(2.7 \pm 0.5) \times 10^{-12} \mu_B$ cm$^2$ A$^{-1}$, is of the same order of magnitude as the values reported by Stamm et al. [8] for Pt, i.e. $5 \times 10^{-12}$ $\mu_B$ cm$^2$ A$^{-1}$ for the topmost atomic layer (and $2 \times 10^{-12}$ $\mu_B$ cm$^2$ A$^{-1}$ for the upper half of the sample). In the cited work the spin diffusion length was determined as $\lambda_{sf}$ (Pt) = 11.4 nm from the thickness dependence of the optical signal (Kerr rotation) and the SHA as $\alpha_{Pt}$ = 0.08, for 300 K and 18-27 μΩcm resistivity. The authors also give the expression which links the induced magnetic moment with the SHA, using the spin drift diffusion model of Zhang [28] within a Boltzmann transport equation framework. In order to calculate the spin Hall angle α for $Cu_{95}Bi_5$ from the measured magnetic moment, one needs the knowledge of further material parameters (see Appendix G). These additional parameters are the spin diffusion length $\lambda_{sf}$, the resistivity, $\rho$, the

Stoner enhancement factor, $F$, and the density of states at the Fermi level, $D(\epsilon_F)$. In our experiment no significant variation of the SHE signal was found within the explored thickness range, i.e. from 20 nm up to 50 nm, as visible in Figure 4. Therefore we lack direct information about the spin diffusion length $\lambda_{sf}$ for our samples. Whereas the resistivity $\rho$ has been determined experimentally as 13 µΩ cm, for $F$ and $D(\epsilon_F)$ we use the values for copper as approximation. As detailed in Appendix G, taking into account the spin accumulation depth profile and the PEEM depth dependent sensitivity, then the implied SHA for any assumed value of $\lambda_{sf}$ can be calculated from the measured magnetic moment (see Figure 11). Depending on the real value of $\lambda_{sf}$, the same magnetic moment would correspond to a different value for SHA for $Cu_{95}Bi_5$. However, there is a lower bound around $\alpha_{CuBi}$ ≈ -0.25 when considering the thicker samples (35, 40 and 50 nm) and a large spin diffusion length around or above 15 nm. In comparison, Niimi et al [15], report a SHA $\alpha_{CuBi}$ = - 0.11 for the CuBi alloy (with the intrinsic limit -0.24 for the skew scattering mechanism) and $\lambda_{sf}$ = 30-50 nm for CuBi samples with lower doping but similar resistivity, measured in non-local spin valves at 10 K. These numbers are compatible with our measurements at 220 K for the thicker samples (35-50 nm), if the spin Hall angle is close to the intrinsic limit of Bi impurities in Cu. However, as can be seen in Figure 4, also the measured magnetic moment for the 20 nm thick sample does not deviate significantly from the others. According to Figure 11, the full magnetic moment for the 20 nm thin sample at the same SHA as the thicker samples would only be expected for the combination of a smaller spin diffusion length, $\lambda_{sf}$ < 8 nm, and a larger SHA $\alpha_{CuBi}$ ≤ - 0.5. The obvious caveats to this conclusion is that it is based on a single data point and using the density of states and Stoner factor of pure Cu. In any case, our results clearly indicate a large spin Hall angle in $Cu_{95}Bi_5$, most likely larger than previously reported.

**Conclusion:**

In summary, using a photoemission electron microscope, we have detected a small XMCD signal at the Cu $L_{3,2}$ absorption edges from the surface of a CuBi electrode, through which an electrical current is flowing. The sign of this signal inverts with the sense of the injected current and is opposite for the $L_3$ and $L_2$ peaks. The size is similar at both peaks with $(10 \pm 2) \times 10^{-12}$ and $(-9 \pm 1) \times 10^{-12}$ cm$^2$ A$^{-1}$ respectively. The XMCD peak shape and position are characteristic of spin accumulation at the Fermi level [21]. The spin accumulation at the $Cu_{95}Bi_5$ surface is ascribed to the spin Hall effect (SHE), reproducing the negative sign of the spin Hall angle of CuBi. The corresponding induced magnetic moment in $Cu_{95}Bi_5$ is $(2.7 \pm 0.5) \times 10^{-12}$ µ$_B$ cm$^2$ A$^{-1}$ per Cu atom as weighted average over the information depth (about 5 nm). The results indicate that the spin Hall angle in CuBi may actually be even larger than previously reported [15]. This large spin Hall efficiency reiterates the potential of CuBi for spin-charge conversion. Furthermore, our results constitute the proof of concept for the direct, interface free and element-selective measurement of the SHE in a single material by means of X-ray spectromicroscopy.

**Acknowledgements:**

We thank Roopali Kukreja for answering our questions about Ref [21] and Sergio O. Valenzuela and Olivier Fruchart for critical reading of the manuscript. We acknowledge M. R. Osorio and D. Granados at IMDEA-nanoscience nanofabrication center for their help in the lithography process.This work has been partially funded by the Spanish Agencia Estatal de Investigación through projects FIS2016-78591-C3-1-R, MAT2017-87072-C4-2- P, RTI2018-097895-B-C42, and RTI2018-095303-B-C53 and by the Comunidad de Madrid through Project NANOMAGCOST-CM P2018/NMT-4321. IMDEA Nanociencia acknowledges support from the Severo Ochoa Programme for Centres of Excellence in R&D (Grant SEV-2016-0686). The work has been supported by the ALBA in-house research program. We thank the Spanish National Center of Electron Microscopy for Scanning Electron Microscopy measurements,

**Appendix A: Details of e-beam lithography**

To define the structures, a bi-layer of MMA 6% in Ethyl-lactate and of 950k molecular weight PMMA 4% in Ethyl-lactate was spin-casted on top of the Si wafer. For both layers, the spin-casting was performed at 3000 rpm for 60 seconds, followed by a soft bake of the resist at 175º C for 60 seconds. The resist was then exposed at a dose of 1000 µC/cm2 with a 100 keV electron beam using a Vistec EBPG 5000Plus electron beam writer. The exposed patterns were then developed by immersion in a solution of methyl-isobutyl-ketone 1:3 in isopropanol (volume) for 60 seconds, followed by immersion in pure isopropanol for an additional 60 seconds. The quality of the developed structures was verified by optical microscopy. Lift-off was performed by immersion in pure acetone.

**Appendix B: Bi segregation after excessive current injection**

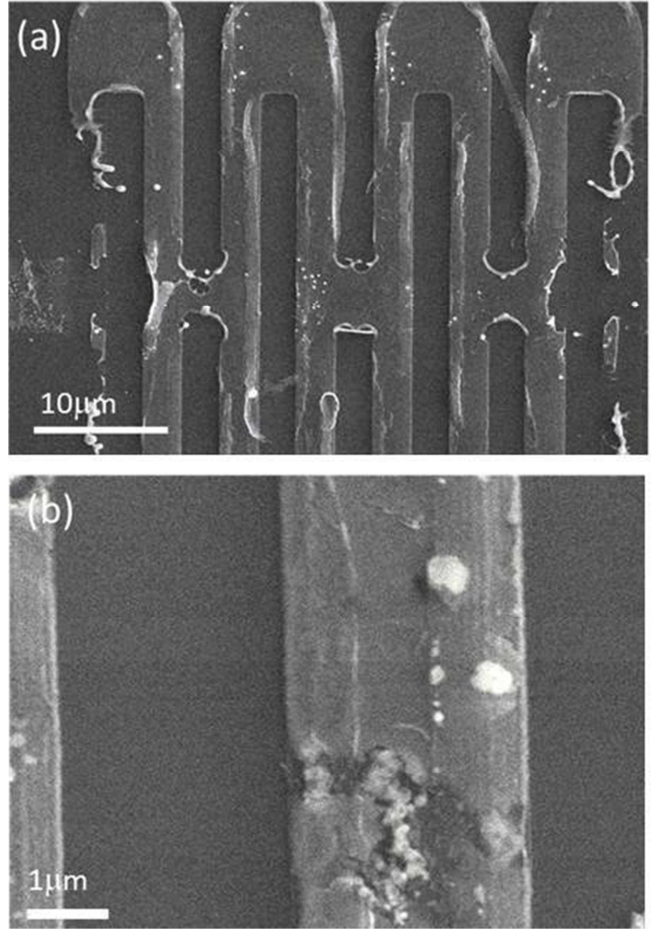

*Figure 5: Scanning electron microscopy images (SEM) of a CuBi electrode structure after damage occurred by current injection of 8 x 10$^7$ A/cm$^2$ at room temperature, exceeding the measurement conditions (max 5.4 x 10$^7$ A/cm$^2$ at 220 K). The white dots were confirmed to be Bi rich.*

**Appendix C. Finite element simulation of the electric and magnetic field**

Figure 6 shows a finite element simulation of a Cu$_{95}$Bi$_5$ electrode of 50 nm thickness and a simulated current density of 3.4 x 10$^6$ A/cm$^2$ assuming a resistivity of 15 µΩcm. As shown in panel (a), the voltage drop along the structure resembles the colormap obtained experimentally (Figure 1(e)). In the experiment, a voltage drop of about 2 V over 2 loops is observed for an applied current density of 1.7 x 10$^7$ A/cm$^2$, while the simulations show 0.93 V over 4 loops. A comparison of these two values then yields a total resistivity of our sample of 13 µΩcm at 220 K, which is expected to include also temperature dependent (phonon) contributions. Note that in Ref. [15], for residual resistivity contribution of the Bi impurities alone (i.e. after subtracting the Cu resistivity) of 3-5 µΩcm at 10 K, a large spin Hall angle is found.

Figure 6(b) shows that the magnitude of the current density in the straight vertical areas of the electrodes, which are used to extract the XMCD signal, is constant.

Figure 6(c) shows the transverse magnetic field component obtained in the upper surface of the electrode, which is about 1.2 mT. Considering the highest current density in our experiments of $j = 7 \times 10^7$ A/cm$^{-2}$, the maximum magnetic field at the surface is then around $B_x \approx 25$ mT. As estimated in Appendix F, the induced paramagnetic moment due to this Oersted field is below $10^{-6}$ $\mu_B$/atom, much smaller than the signal that we measure.

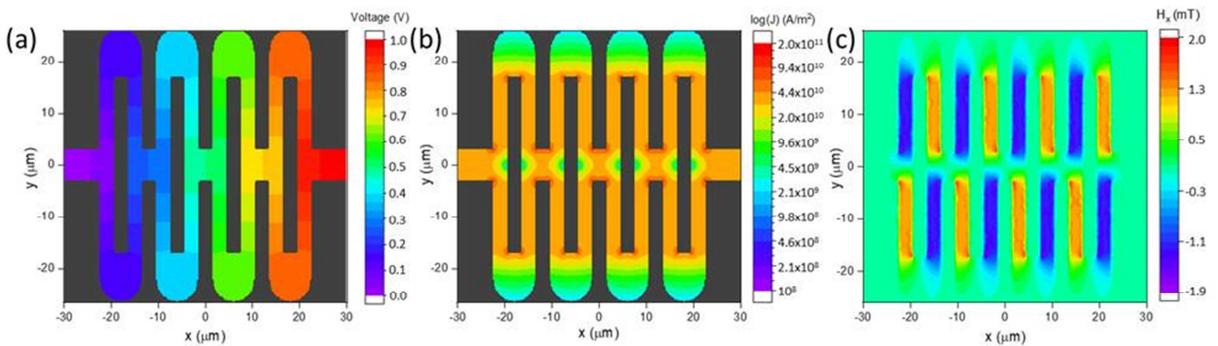

*Figure 6: Finite element simulation of the CuBi electrodes with current flow: (a) voltage drop along the structure (b) current density along the structure in logarithmic scale and (c) transverse magnetic field component.*

**Appendix D: X-ray Photoelectron spectra (XPS) of oxidized samples**

Figure 7 shows the Bi 4f X-ray Photoelectron spectra (XPS) of a sample from which the capping layer has been removed by Ar sputtering, before and after exposition to air. The spectrum before exposition shows mostly metallic Bi signal with a small oxide component while the sample exposed to air after the capping removal showed almost fully oxidized Bi.

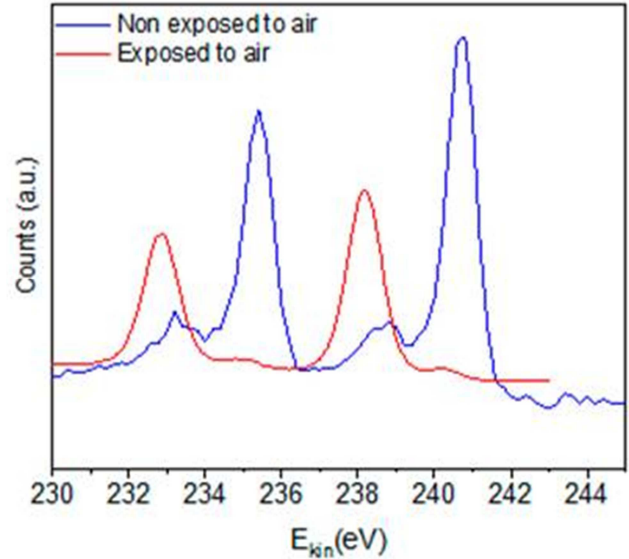

*Figure 7: Photoelectron spectrum of the Bi 4f core level of a sample after capping layer removal, before and after exposition to air.*

**Appendix E: Data treatment**

a) **Fit process to obtain SHE raw signal**

To extract the spin accumulation due to the SHE, the pixel averaged XMCD signal coming from each region marked with a yellow square in Figure 8(a) was obtained from XMCD images as shown in Figure 8(b). The contrast scale (black to white) in panel (b) is rather large, around 2.1% because it needs to accommodate the background. The insulating silicon oxide surface appears in a different shade due to charging effects. In order to reduce the background contribution coming from the voltage drop along the structure and the difference between the beam intensity for the two polarizations, each single XMCD value indicated as data point in Figures 8 (c) - (f) is the difference (asymmetry) of the signal from two vertically aligned regions with the same potential but opposite currents, e.g. labelled 1a and 1b in panel (a).

As can be seen in Figures 8(c)-(f), where the signal is plotted vs the position (electrode branch number), the background is then strongly reduced but still present. The signal of the spin Hall effect is a small alternating signal on top of the background. To determine it, a parabolic fit (see inset) including an oscillating term S was performed for all the measurements. The fit parameter S is then twice the XMCD signal coming from the spin accumulation due to the spin Hall effect in a single area. For panels (c) -(f), we have selected on purpose examples with little and large background variations.

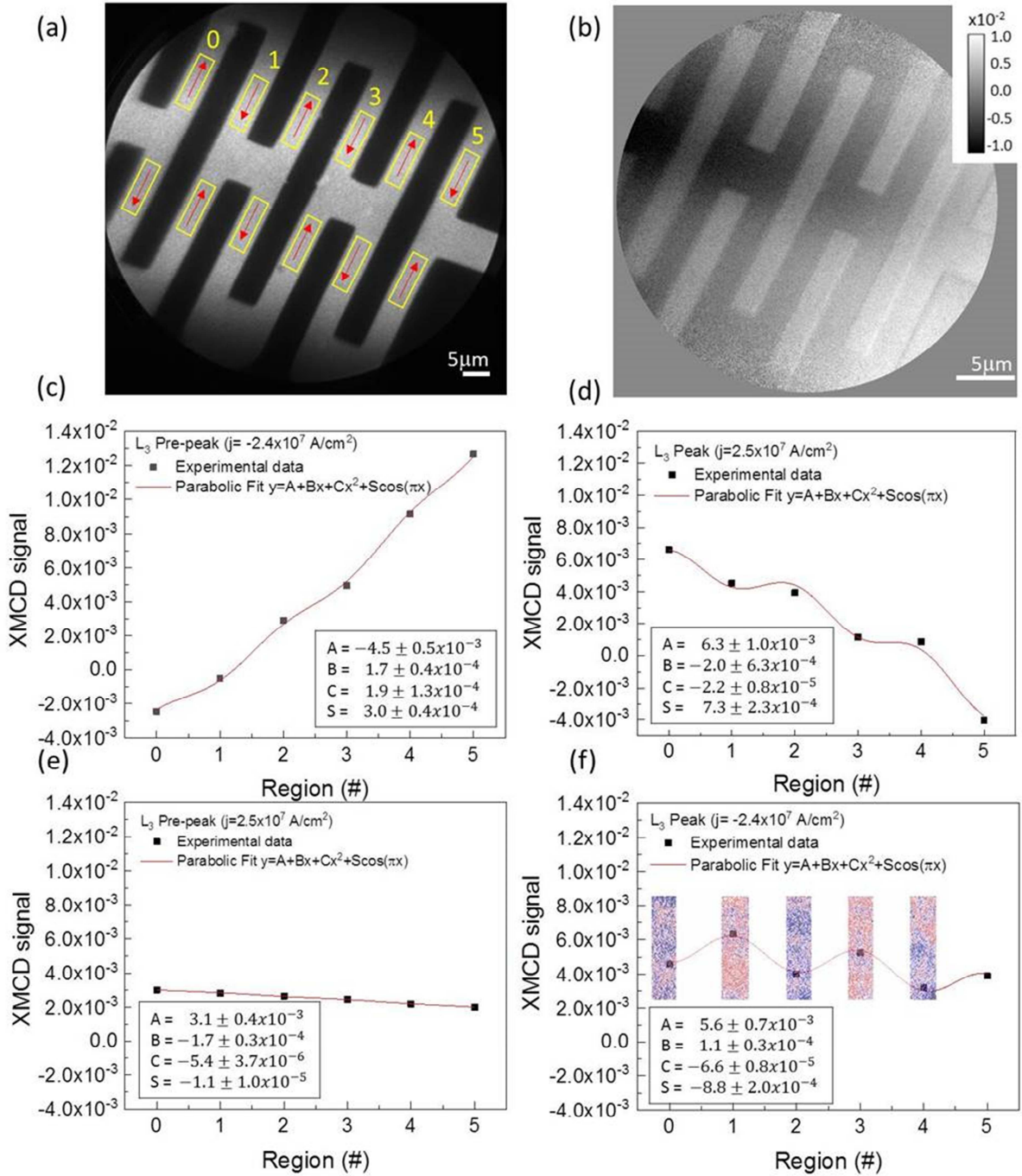

Figure 8: (a) XAS image together with the regions (yellow squares) used to extract the XMCD signal. The red arrows show the direction of the current flow. (b) Example of an XMCD image at the $L_3$ peak photon energy. (c) –((f) Examples of XMCD values as difference of the upper and lower area vs branch number of the electrode for the $L_3$ pre-peak and $L_3$ peak photon energy. A parabolic fit (inset) is used to compensate the background and extract the alternating SHE contribution "S". We have chosen a pair of examples with very low and very high background variation.

**b) Pre-peak subtraction**

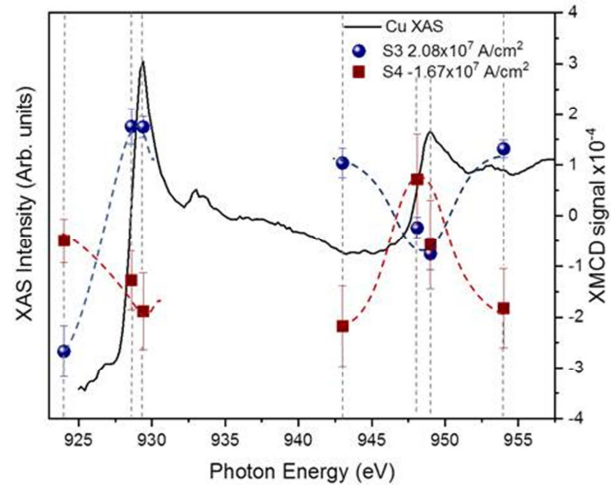

*Figure 9: X-ray absorption spectrum of the Cu L edge together with the value of the XMCD signal obtained at different energies as the average of at least 3 measurements. Dashed lines are guide to the eye.*

Plotting the value of the S fit parameter (S/2 in Figure 9) as a function of the energy one sees that the XMCD signal has opposite values in $L_3$ and $L_2$ edges. However, even for off-peak energies where no XMCD signal is expected, an offset is typically found. As can be seen in Figure 9, this offset can be different for the different measurement series. For example, while raw data at the $L_2$ peak is very similar for opposite current directions, there is a clear opposite signal when comparing values at $L_2$ with the corresponding pre-peak and after peak measurements. The offset was found to vary little over time and photon energy within a given data set, recorded at constant current and within several hours. Therefore a normalization of the XMCD signal for the $L_3$ and $L_2$ peaks as the difference of the value obtained at the absorption peak and the value obtained by an immediately preceding measurement at a pre-peak photon energy was performed. The analysis of all data where this method was applicable, from four different samples and different measurement campaigns, is fully consistent and shown in Figure 4.

In Figure 10, we show an example to assess the influence of the pre-peak subtraction. Black symbols represent values obtained directly for a single photon energy ($L_3$ peak), averaged over several images. However, only for two current densities the pre-peak measurement is available (red symbols, after subtraction) and only those were included in Figure 3(c). The dashed line is the same linear fit to the data as in Figure 3(c) and it also describes these measurements reasonably. Although in this case the influence of the offset is only moderate, we included only those measurements into the final analysis where the offset correction could be performed.

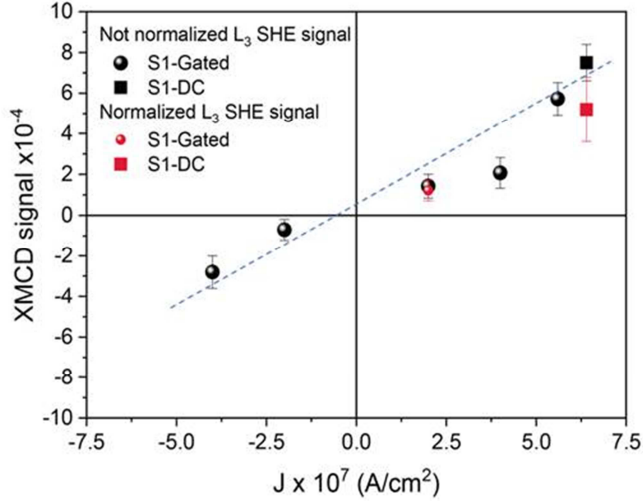

*Figure 10: Comparison of XMCD values at the $L_3$ peak for a single sample, with and without offset subtraction (illustrating a case where not all pre-peak measurements were measured). Black symbols denote data without subtraction of the pre-peak offset, red symbols with the offset subtracted. The dashed line is the same as in Figure 3 (c).*

### Appendix F. Paramagnetic moment from Oersted field

One needs to consider the effect of the current-induced Oersted field, which also produces a small paramagnetic spin polarization at the sample surface: using m = $\mu_B^2$ D($E_F$) $\mu_0$ H [29] where D($E_F$) is the density of states at the Fermi Energy (taking around 0.55 eV$^{-1}$ per atom [30]), the expected moment for Cu would be m = 3.2 x 10$^{-5}$ $\mu_B$ T$^{-1}$ per atom. Following Reference [31] we estimate the transverse magnetic field component at the top surface as $B_x$ = $\mu_0$ j z, where j is the current density and z the vertical distance from the stripe center. The highest value in our experiments is thus (for j = 7 x 10$^7$ A cm$^{-2}$ and z = 25 nm) $B_x$ ≈ 22 mT (in agreement with the simulated value 25 mT in Appendix C). Therefore, the paramagnetic moment resulting from the Oersted field (below 10$^{-6}$ $\mu_B$ /atom) is much smaller than the signal that we measure.

### Appendix G. Analysis of the magnetic moment

According to Stamm et al. [8] (combining equations S5 and S6), based on the work of Zhang [28], the spin accumulation per current density j, at depth z, is given by:

$$M_j(z) = \frac{M_y(z)}{j} = \lambda_{sf} \alpha \rho \frac{\sinh\left(\frac{t-2z}{2\lambda_{sf}}\right)}{\cosh\left(\frac{t}{2\lambda_{sf}}\right)} D(\epsilon_f) F \qquad (1)$$

in units of $\mu_B$ per atom. Here $t$ is the sample thickness, $\lambda_{sf}$ is the spin diffusion length, α is the spin Hall angle, $\rho$ the resistivity, $D(\epsilon_f)$ is the total density of states at the Fermi level and $F$ is the Stoner enhacement factor. This expression is valid for $\lambda_{sf}$ reasonably smaller than t, and thus as will be seen later, for the parameter range for which we can make some statement.

The PEEM depth dependent sensitivity is modeled by the following function:

$$H(z) = e^{-(z/d)} \quad (2)$$

With d = 2 nm [23].

The PEEM detected moment per atom and unit of current density is then:

$$m_j(t) = \int_0^t dz\, M_j(z) H(z) \Big/ \int_0^t dz\, H(z) \quad (3)$$

Solving the integrals and substituting values for $\rho$, $D(\epsilon_f)$ and $F$, we obtain an expression for α as function of $\lambda_{sf}$ and the sample thickness t in nm.

$$8.0366 \times 10^{-8} \alpha \lambda_{sf}^2 \frac{e^{-\frac{t}{2\lambda_{sf}}}}{2\cosh\left(\frac{t}{2\lambda_{sf}}\right)\left(e^{\frac{t}{d}}-1\right)} \left[ \frac{e^{t\left(\frac{1}{\lambda_{sf}}+\frac{1}{d}\right)}-1}{\lambda_{sf}+d} + \frac{e^{\frac{t}{\lambda_{sf}}}-e^{\frac{t}{d}}}{\lambda_{sf}-d} \right] = 2.7 \times 10^{-12} \quad (4)$$

The prefactors used are $D(\epsilon_f) = 0.55$ eV$^{-1}$ per atom [31], and the Stoner enhancement factor F is given by:

$$F = \frac{1}{1-In(\epsilon_f)} = 1.124 \quad (5)$$

Since $In(\epsilon_f) = 0.11$ according to Ref. [32]. The values for these two parameters are thus taken for Cu as an approximation. A more precise determination of these parameters for Cu$_{95}$Bi$_5$ is beyond the scope of this work, which is primarily dedicated to the demonstration of a measurement of the SHE with X-ray spectroscopy, but it remains interesting for the future. The experimental resistivity of the samples is 13 $\mu\Omega cm$. Equation (4) yields a function $\alpha(\lambda_{sf})$ based on the determined magnetic moment from XMCD-PEEM. It further depends on the sample thickness t, as the spin accumulation is reduced when the thickness is comparable to the spin diffusion length $\lambda_{sf}$. Since there is no direct information on the spin diffusion length $\lambda_{sf}$ from our data, the equation is to be understood as: if the real spin diffusion length was $\lambda_{sf}$, then the spin hall angle which corresponds to the magnetic moment would be α.

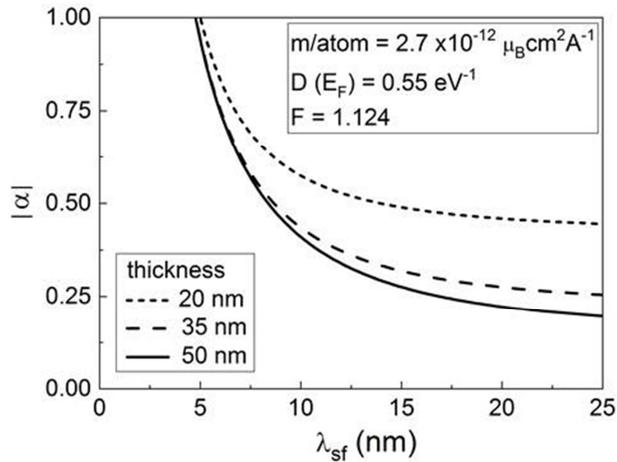

*Figure 11: Calculated relation of the spin diffusion length and the spin Hall angle in $Cu_{95}Bi_5$ for the magnetic moment measured by XMCD-PEEM (linear fit in Figure 4). Further calculation parameters are the measured resistivity and the density of states at Fermi level and Stoner factor taken for Cu. Curves of equal magnetic moment are plotted for different sample thickness. For the thinnest sample higher values for the spin Hall angle are required to match experimental observations.*

Figure 11 shows the curves for the same magnetic moment and 3 characteristic thicknesses in our study, namely: 20, 35 and 50 nm. The curves for the 35 and 50 nm thick samples indicate a lower bound of the SHA $|\alpha| \approx 0.25$ for long spin diffusion length $\lambda_{sf}$ around or above 15 nm; those values are compatible with the results of Niimi et al. [15]. However if the real value of $\lambda_{sf}$ is lower, then the absolute of the SHA $|\alpha|$ needs to be higher in order to be still compatible with the measured magnetic moment. The single data point available for the 20 nm thick sample points into this direction, because no significant reduction of the measured XMCD signal was found (see Figure 4). According to the curve plotted in Figure 11, in order to find a comparable moment in such a thin sample as in the thicker ones, the spin diffusion length would have to be smaller, less than about 8 nm, and the SHA much larger $|\alpha| > 0.5$. While it should be noted that this treatment is at the limit of the applicability of the spin drift diffusion model (t could be comparable to $\lambda_{sf}$), clearly further experiments are needed to gain more insights into the SHA of $Cu_{95}Bi_5$.

**References:**

[1] J. E. Hirsch, Spin Hall Effect, Phys. Rev. Lett. 83, 1834 (1999)